\documentclass[conference]{IEEEtran}
\usepackage{amsmath,amssymb}

\usepackage{graphicx}
\usepackage{slashbox}

\ifCLASSINFOpdf
\else
\fi
\begin{document}

\title{A Novel Scattered Pilot Design for FBMC/OQAM Systems }

%\author{\IEEEauthorblockN{Pengfei Sun}
%\IEEEauthorblockA{School of Electrical and\\Computer Engineering\\
%Georgia Institute of Technology\\
%Atlanta, Georgia 30332--0250\\
%Email: http://www.michaelshell.org/contact.html}
%\and
%\IEEEauthorblockN{Homer Simpson}
%\IEEEauthorblockA{Twentieth Century Fox\\
%\and
%\IEEEauthorblockN{James Kirk\\ and Montgomery Scott}
%\IEEEauthorblockA{Starfleet Academy\\
%San Francisco, California 96678-2391\\


\author{\IEEEauthorblockN{Pengfei Sun\IEEEauthorrefmark{1},
Fang Yuan\IEEEauthorrefmark{1},
Bin Yu\IEEEauthorrefmark{1},
Dalin Zhu\IEEEauthorrefmark{1}}
\IEEEauthorblockA{\IEEEauthorrefmark{1}Beijing Samsung Telecom R\&D Center\\
Email:pf.sun@samsung.com, Yuanfang@BUAA.com, bin82.yu@samsung.com, dalin.zhu@samsung.com}
\IEEEauthorblockA{\IEEEauthorrefmark{1}18F TaiYangGong Plaza, Chaoyang District, Beijing, 100028, China}
}


\maketitle

\begin{abstract}
%\boldmath
Filter bank multi-carrier with offset quadrature amplitude modulation (FBMC/OQAM) has been heavily studied as an alternative waveform for 5G systems. Its advantages of higher spectrum efficiency, localized frequency response and insensitivity to synchronization errors may enable promising performance when orthogonal frequency division multiplexing (OFDM) fails. However, performing channel estimation under the intrinsic interference has been a fundamental obstacle towards adopting FBMC/OQMA in a practical system. Several schemes are available but the performance is far from satisfaction. In this paper, we will show the existing methods are trapped by the paradigm that a clean pilot is mandatory so as to explicitly carry a reference symbol to the receiver for the purpose of channel estimation. By breaking this paradigm, a novel dual dependent pilot scheme is proposed, which gives up the independent pilot and derives dual pilots from the imposed interference. By doing this, the interference between pilots can be fully utilized. Consequentially, the new scheme significantly outperforms existing solutions and the simulation results show FBMC/OQAM can achieve close-to-OFDM performance in a practical system even with the presence of strong intrinsic interference.
\end{abstract}


% \fi
\IEEEpeerreviewmaketitle

\section{Introduction}
Nowadays, both industry and academia are working towards the 5G mobile communication systems. Supporting versatile wireless services with a universal technology are becoming the consensus among researchers. These service driven requirements including ultra low latency access, fragmented spectrum utilization, robustness in super high speed etc. have raised unprecedented challenges for the future mobile communication technology. As future wireless communication scenarios always involve a very harsh communication environment, it has been questioned if OFDM is still the best choice for all cases. The recent research trend has been looking into other alternative modulation schemes trying to find a concrete answer \cite{METIS}.

Among these alternative modulation schemes, filter bank multi-carrier with offset quadrature amplitude modulation (FBMC/OQAM) is one of the competitive candidates for 5G scenarios. Due to the absence of cyclic prefix (CP), FBMC/OQAM achieves very high spectrum efficiency. The properly designed pulse shaping filter enables FBMC/OQAM a more localized frequency response, which leads to a lower out-of-band leakage and enhanced robustness to synchronization errors. By a careful system design, these advantages may enable promising performances in certain scenarios while OFDM fails \cite{FBMC_Tutorial}.
However, the channel estimation in FBMC/OQAM system has been a fundamental issue for a long time. FBMC/OQAM only sustains a real field orthogonality which means imaginary interference is imposed to each subcarrier. Such intrinsic interference can be handled by using real valued pulse amplitude modulation (PAM) for data symbols. However, this intrinsic interference severely damages the pilot signal in the channel estimation stage leading to poor estimation accuracy. A degraded channel estimation performance means all the aforementioned advantages of FBMC/OQAM is not guaranteed. Thus, channel estimation of FBMC/OQAM is a fundamental obstacle to the practical adoption of FBMC/OQAM.
Several schemes are available for solving the channel estimation problem in FBMC/OQAM, which can be categorized by either preamble based or scattered pilot based solutions. In this paper we concentrate on scattered pilot based solutions. The most straightforward method is to mute the pilot's surrounding symbols to avoid interference. However, such method requires unacceptable overhead in a practical system. Auxiliary pilot scheme (AUP) has been proposed in \cite{AUP}, which applies an auxiliary pilot to neutralize the interference on the pilot subcarrier. Composite pilot pair (CPP) utilizes two auxiliary pilots to achieve the same goal \cite{CPP}. AUP and CPP apply the same principle: utilize auxiliary pilots to generate an inverse interference to the primary pilot. Such principle promises a clean primary pilot to the receiver while the overhead is reduced. However, the primary pilot constraint requested by this principle will result in a low efficient channel estimation, which will be discussed throughout this paper. A novel scheme is proposed in this paper, which breaks the paradigm of the primary pilot constraint and results in significantly improved performance.

The rest of the paper is organized as: section II introduces the system model. The novel scheme is proposed in section III where several existing schemes are jointly discussed for analogy. Section IV shows the simulation results and section V concludes the paper.
\section{System Model}\label{SystemModel}
The baseband model of FBMC/OQAM signal can be written as:
\begin{equation}
s[k] = \sum_{n\in\mathbb{Z}}\sum_{m=1}^{M-1}a_{m,n}\underbrace{g\left[m-nM/2\right]j_{m,n}e^{j\frac{2\pi}{M}km}}_{g_{m,n}[k]}
\end{equation}
where $a_{m,n}$ is the PAM symbol modulated to each subcarrier. For simplicity, we use index pair $(m,n)$ to denote the $m^{th}$ subcarrier of the $n^{th}$ symbol. Denote $\rho^{2}=E(a_{m,n}^{2})\  \forall (m,n)$ as the power of the PAM symbol. $M$ is the total number of subcarriers of the system. $g[k]\ (k=0,1,...KM-1)$ is the prototype filter and $K$ is the overlapping factor. In addition, $j_{m,n}$ is the phase factor alternating between $1$ and $j$.
$g_{m,n}[k]$ is denoted as the synthesis filter used for modulating the $(m,n)^{th}$ symbol.
At the receiver, the received signal is passed to a bank of analysis filter. Assuming no channel distortion and additive noise,  the output at the $(m,n)^{th}$ analysis filter is:
\begin{eqnarray}\label{Rec_Output}
r_{m,n}&=&\sum_{k=-\infty}^{+\infty}s[k]g^{*}_{m,n}[k]\nonumber\\
&=&\sum_{k=-\infty}^{+\infty}\sum_{m'=0}^{M-1}\sum_{n'\in\mathbb{Z}}a_{m',n'}g_{m',n'}[k]g^{*}_{m,n}[k]
\end{eqnarray}
Denote:
\begin{equation}\label{Amb_Func}
\Lambda_{m',n',m,n}=\sum_{k=-\infty}^{+\infty}g_{m',n'}[k]g^{*}_{m,n}[k]
\end{equation}
as the ambiguity function, the perfect reconstruction (PR) property of FBMC/OQAM can be written as:
\begin{equation}
\Lambda_{m',n',m,n}=\left\{
\begin{aligned}
  &1\ \ \  \text{ if } (m,n)=(m',n') \\
&j\beta_{m',n',m,n}\ \   \text{ otherwise} \\
\end{aligned}
\right.
\end{equation}
where $\beta_{m',n',m,n}$ is the real valued interference coefficient. Consequentially, (\ref{Rec_Output}) can be written as:
\begin{equation}
r_{m,n}=a_{m,n}+\sum_{(m',n')\in \mathbf{\theta}_{m,n}}j\beta_{m',n',m,n}a_{m',n'}
\end{equation}
where $\mathbb{\theta}_{m,n}$ is the index set of subcarriers surrounding the $(m,n)^{th}$ subcarrier. Consider the signal is passing through a multi-path channel with additive noise and assume the channel is flat over several neighboring subcarriers, the received signal could be modeled as:
\begin{equation}
r_{m,n}=Ha_{m,n}+H\sum_{(m',n')\in \theta_{m,n}}\beta_{m',n',m,n}a_{m',n'}j+n_{m,n}
\end{equation}
where $H$ is the channel frequency response. Consider a pilot symbol $p_{m,n}$ is inserted within a data block for channel estimation: $a_{m,n}=p_{m,n}$, then the received pilot becomes:
\begin{equation}\label{Pil_Sig}
r_{m,n}=Hp_{m,n}+\underbrace{H\sum_{(m',n') \in \theta_{m,n}}\beta_{m',n',m,n}a_{m',n'}}_{I_{m,n}}+n_{m,n}
\end{equation}
From (\ref{Pil_Sig}), the received pilot is subjective to the interference $I_{m,n}$ which arises from surrounding data symbols in addition to the additive noise.
\section{Dual Dependent Pilot Scheme}
In this section, we present a novel dual dependent pilot scheme (DDP) to solve the problem formulated in section \ref{SystemModel}.
\subsection{Brief review}
Auxiliary pilot scheme (AUP) is a well known scheme that neutralizes the interference term formulated in (\ref{Pil_Sig}) by using an auxiliary pilot. Denote $p_{m,n}$ as the prime pilot, an auxiliary pilot $u_{\check{m},\check{n}}$ is inserted at the nearby subcarrier $(\check{m},\check{n})$. Then, the received prime pilot can be described as:
\begin{eqnarray}\label{AUP_Mod}
\begin{split}
r_{m,n}=&Hp_{m,n}+H(\beta_{\check{m},\check{n},m,n}u_{\check{m},\check{n}}+\\
&\sum_{\begin{subarray}{1}(m',n') \in \theta_{m,n}\setminus(\check{m},\check{n})\end{subarray}}\beta_{m',n',m,n}a_{m',n'})j+
n_{m,n}
\end{split}
\end{eqnarray}
In such way, the auxiliary pilot can be calculated to neutralize the interference:
\begin{eqnarray}\label{AUP_Sol}
u_{\check{m},\check{n}}=-\frac{\sum_{\begin{subarray}{1}(m',n') \in \theta_{m,n}\\(m',n')\not=(\check{m},\check{n})\end{subarray}}\beta_{m',n',m,n}a_{m',n'}}{\beta_{\check{m},\check{n},m,n}}
\end{eqnarray}
As a result, as long as $p_{m,n}$ equals to a pre-defined value, conventional channel estimation can be performed with zero interference. Pre-defined value means the prime pilot is independent to any other symbols. On the other hand, the auxiliary pilot $u_{\check{m},\check{n}}$ is calculated according to the surrounding data symbols, so we deem it as a "Dependent Pilot".
Another scheme named composite pilot pair (CPP) applies a similar principle, three pilots in total are placed close to each other. The central pilot is a prime pilot that is independent to other symbols. The other two pilots are generated to neutralize the interference suffered by the prime pilot. The two auxiliary pilots are calculated by:
\begin{eqnarray}
\begin{split}
c_{m,n-1}=\frac{1}{\sqrt{2}}a^{*}-\frac{1}{2}\sum_{\begin{subarray}{1}(m',n') \in \theta_{m,n}\setminus(m,n\pm1)\end{subarray}}\frac{\beta_{m',n',m,n}a_{m',n'}}{\beta_{m,n-1,m,n}}\\
c_{m,n+1}=-\frac{1}{\sqrt{2}}a^{*}-\frac{1}{2}\sum_{\begin{subarray}{1}(m',n') \in \theta_{m,n}\setminus(m,n\pm1)\end{subarray}}\frac{\beta_{m',n',m,n}a_{m',n'}}{\beta_{m,n+1,m,n}}
\end{split}
\end{eqnarray}
where $a^{*}$ is a data symbol superimposed on these two auxiliary pilots to reduce the overhead.
In summary, the common principle can be found in both schemes: one prime pilot, whose value is independent to other symbols, is required to explicitly carry a pre-defined reference symbol to assist receiver's channel estimation. Auxiliary pilots, on the other hand, are dependent to surrounding data symbols so that they could generate an inverse interference to the prime pilot.
\subsection{Novel dual dependent pilot scheme}
\begin{figure}
\centering
\includegraphics[width=6cm]{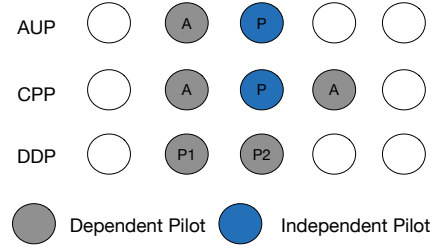}
\caption{Structure of dependent and independent pilots for different schemes}\label{Figure1}
\end{figure}

In this subsection, we propose a dual dependent pilot (DDP) scheme, which applies a totally different principle compared to above schemes. For simplicity, we assume that two real valued pilots $p_{m,n}^{1}$, $p_{m,n+1}^{2}$ are inserted at subcarriers $(m,n)$ and
$(m,n+1)$ respectively as a specific case. At the receiver, the received signal can be written as:
\begin{equation}\label{Dual_Pilot}
\begin{split}
r_{m,n}=&Hp_{m,n}^{1}+Hi_{1}j\\&+H\beta_{m,n+1,m,n}p_{m,n+1}^{2}j+n_{m,n}\\
r_{m,n+1}=&Hp_{m,n+1}^{2}+Hi_{2}j\\&+H\beta_{m,n,m,n+1}p_{m,n}^{1}j+n_{m,n+1}
\end{split}
\end{equation}
where
\begin{equation}
\begin{split}
i_{1}&=\sum_{(m',n')\in \mathbf{\theta}_{m,n}\setminus (m,n+1)}\beta_{m',n',m,n}a_{m',n'}\\
i_{2}&=\sum_{(m',n')\in \mathbf{\theta}_{m,n+1}\setminus(m,n)}\beta_{m',n',m,n+1}a_{m',n'}
\end{split}
\end{equation}
Observing (\ref{Dual_Pilot}), the receiver has two received pilots in complex domain. The transmitted pilots align in the real domain and the interferences align in the imaginary domain. That is, the received pilots have four degrees of freedom while pilots and interferences have two degrees of freedom respectively. If receiver performs linear processing on the received pilots to reduce their degree of freedom, it is possible to manipulate the transmitted pilot so that the interference term is neutralized after the processing. Define the aforementioned linear processing on the received pilot as:
\begin{eqnarray}\label{Lin_Proc}
r= r_{m,n}+\alpha r_{m,n+1}
\end{eqnarray}
and a clean pilot is pre-defined as $x+yj$. In such framework, the interference neutralization principle becomes: calculate the weight factor $\alpha$ and transmitted pilots $p_{m,n}^{1}, p_{m,n+1}^{2}$ so that:
\begin{equation}\label{Rec_Clean}
\begin{split}
r= x+yj+n_{m,n}+\alpha n_{m,n+1}
\end{split}
\end{equation}
which is equivalent to:
\begin{equation}\label{Int_Neu}
\begin{split}
&p_{m,n}^{1}+i_{1}j+\beta_{m,n+1,m,n}p_{m,n+1}^{2}j+\\
&\alpha (p_{m,n+1}^{2}+i_{2}j+\beta_{m,n,m,n+1}p_{m,n}^{1}j)=x+yj
\end{split}
\end{equation}
For any given $(x,y)$, there are numerous solutions of $(\alpha, p_{m,n}^{1}, p_{m,n+1}^{2})$ to equation (\ref{Int_Neu}). If we select $\alpha=0$ and let $y=0$, the transmitted pilot $(p_{m,n}^{1}, p_{m,n+1}^{2})$ based on (\ref{Int_Neu}) can be calculated:
\begin{equation}
\begin{split}\label{AUP_2Form}
p_{m,n}^{1}=x,\ \
p_{m,n+1}^{2}=-\frac{i_{1}}{\beta_{m,n+1,m,n}}
\end{split}
\end{equation}
Solution in (\ref{AUP_2Form}) is exact the same result as AUP's solution described in (\ref{AUP_Sol}). Therefore, finding a $(\alpha, p_{m,n}^{1}, p_{m,n+1}^{2})$ that satisfies (\ref{Int_Neu}) with a given $(x, y)$ is the general form of solution to the interference neutralization problem. Although all solutions can neutralize the interference and thus promises a clean pilot, the channel estimation performance of each solution varies. We herein discuss another solution derived from the general form of solution. Select $\alpha$ as pure imaginary with unit power: $\alpha= \pm j$ , solving (\ref{Int_Neu}) gives:
\begin{eqnarray}\label{DDP}
p_{m,n}^{1}=\frac{x+i_{2}}{1\mp \beta_{m,n,m,n+1}},
p_{m,n+1}^{2}=\frac{y-i_{1}}{ \beta_{m,n+1,m,n}\pm 1}
\end{eqnarray}
From (\ref{DDP}), it can be found that the new solution requires to calculate two pilot symbols when $\alpha\not=0$ is selected. Each pilot depends on the interference imposed on the other pilot and the desired clean pilot signal is implicitly carried by two pilots. Thus such scheme can be deemed as ``Dual Dependent Pilot (DDP)'' as comparing with the ``Single Dependent Pilot (SDP)'' adopted by AUP. It is obvious that CPP is also another specific solution that can be derived from the general solution where three subcarriers are required. In addition, pilot and data must be superimposed in CPP to counter overhead, which leads to complex receiver algorithm. In summary, sticking to an independent pilot is just a conventional wisdom instead of a proved requisite. As shown above, DDP scheme has following proved features: 1) two PAM symbols are the minimum requirement to generate a clean complex reference symbol without requirement of independent pilots; 2) all the known interference can be utilized; 3) the interference neutralization solution does not rely on specific pilot symbol. Due to the space limitation, we do not present detailed discussion of each feature here. But these features are vital in practical wireless communication systems. A demonstration showing the pilot structure difference between AUP, CPP and DDP scheme can be found in Fig. \ref{Figure1}.
\subsection{Performance Analysis}\label{Perf_Com}
As CPP involves three subcarriers, it is difficult to make a fair comparison with DDP. As aforementioned, there are also infinite solutions to the interference neutralization problem for three subcarriers case, among which CPP may not be the best. In addition, a even number of subcarriers for reference signal is a more acceptable design in practice. Thus we only present performance analysis of AUP and DDP to show the benefit of giving up the independent pilot, which can be extended to more cases. Both AUP and DDP involve dependent pilots, which are subjective to the surrounding random data symbols. Therefore the performance analysis must take into account of the varying power of the dependent pilots. For AUP, based on (\ref{Rec_Clean}) and $\alpha=0$, the received pilot after processing (\ref{Lin_Proc}) is:
 \begin{eqnarray}\label{Rec_AUP}
r_{AUP}=x+n_{m,n}
\end{eqnarray}
If $Var(n_{m,n})=\sigma ^{2}$ and $E(x^{2})=e^{2}$, the SNR for AUP can be derived:
\begin{equation}\label{SNR_AUP}
SNR_{AUP}=\frac{E(x^{2})}{Var(n_{m,n})}=\frac{e^{2}}{\sigma ^{2}}
\end{equation}
For DDP, based on (\ref{Rec_Clean}) and $\alpha=\pm j$ the received pilot after processing is:
 \begin{eqnarray}\label{Rec_DPP}
r_{DDP}=x+yj+n_{m,n}\pm n_{m,n+1}
\end{eqnarray}
If $E(x^{2})=E(y^{2})=e^{2}$ and $Var(n_{m,n})=Var(n_{m,n+1})=\sigma ^{2}$, then, the SNR for DDP can be derived:
\begin{equation}\label{SNR_DDP}
\begin{split}
SNR_{DDP}=&\frac{E(x^{2})+E(y^{2})}{Var(n_{m,n}\pm jn_{m,n+1})}\\
=&\frac{2e^{2}}{2\sigma ^{2}\pm 2Cov(n_{m,n},jn_{m,n+1})}
\end{split}
\end{equation}
According to (\ref{SNR_AUP}), the SNR for clean pilot in AUP only depends on the power of the pre-defined reference symbol and the noise power. However, according to (\ref{SNR_DDP}), the covariance of the noises in two subcarriers also determines the SNR for clean pilot in DDP in addition to the power of pre-defined reference symbol. This covariance is not zero as the noise is dependent between two adjacent subcarriers. The pilots' transmission power required by the two schemes are also different. Denote $P$ as the power required for all pilots:
\begin{equation}\label{Power_Com}
\begin{split}
P_{AUP}=&E[ (p_{m,n}^{1})^{2}+(p_{m,n+1}^{2})^{2}]=e^{2}+\frac{E(i_{1}^{2})}{\beta_{m,n+1,m,n}^{2}}\\
P_{DDP}=&E[ (p_{m,n}^{1})^{2}+(p_{m,n+1}^{2})^{2}]\\
=&\frac{e^{2}+E(i_{2}^{2})}{(1\mp\beta_{m,n+1,m,n})^{2}}+\frac{e^{2}+E(i_{1}^{2})}{(\beta_{m,n+1,m,n}\pm 1)^{2}}
\end{split}
\end{equation}
%where
%\begin{equation}
%\begin{split}
%E(i_{1}^{2})=&\rho^{2}\sum_{(m',n')\in \mathbf{\theta}_{m,n}\setminus (m,n+1)}\beta_{m',n',m,n}^{2}\\
%E(i_{2}^{2})=&\rho^{2}\sum_{(m',n')\in \mathbf{\theta}_{m,n+1}\setminus(m,n)}\beta_{m',n',m,n+1}^{2}
%\end{split}
%\end{equation}
According to (\ref{Amb_Func}), it can be known that $\beta_{m,n+1,m,n}=-\beta_{m,n+1,m,n}$. Thus if $\beta_{m,n+1,m,n}>0$, then we could find the minimum power required for DDP is when $\alpha=-j$:
 \begin{equation}\label{PDPP_Min}
 P_{DDP}^{Min}=\frac{e^{2}+E(i_{1}^{2})+E(i_{2}^{2})}{(1+\beta_{m,n+1,m,n})^{2}}
 \end{equation}
 and vice versa for $\beta_{m,n+1,m,n}<0$.
 According to (\ref{SNR_AUP})(\ref{SNR_DDP})(\ref{Power_Com}) and (\ref{PDPP_Min}), it is nontrivial to determine which one has better performance. A simple comparison is to normalize the received SNR for the clean pilots for both schemes and evaluate the transmission power. We take PHYDYAS filter as an example here \cite{PHYDYAS}, which gives the impulse response as shown in Table \ref{PHY}. The interference coefficient $\beta_{m',n',m,n}$ can be derived from the table, e.g. $\beta_{m,n+1,m,n}=0.5646$. In order to keep the SNR same for both data symbols and clean pilots, we set $E(a^{2})=1/2$ and set $SNR_{AUP}=SNR_{DDP}$, then $E(i_{1}^{2})=E(i_{2}^{2})\approx 0.33$ and $Var(n_{m,n}\pm jn_{m,n+1})=3\sigma^{2}$ can be calculated which leads to $P_{DDP}^{Min}=1.5$ and $P_{AUP}=2.0$.
 According to above analysis, DDP requires less pilot transmission power to achieve a targeted channel estimation performance. From another perspective, DDP will have better channel estimation performance when pilot transmission power is normalized. This gain is because DDP utilizes the mutual interference between two pilots while AUP only utilizes the interference from auxiliary pilot to the primary pilot and wastes the interference from the primary pilot to the auxiliary pilot.
 \begin{table}
 \caption{Impulse response of the prototype filter}\label{PHY}
\begin{tabular}{|c|c|c|c|c|c|c|c|}
\hline
\backslashbox{Sub.\kern-5em}{Time} & n-3 & n-2 & n-1 & n & n+1 & n+2 & n+3 \\
\hline
m-1 & 0.043j & -0.125 & -0.206j & 0.2393 & 0.206j&-0.125 & -0.043j \\
\hline
m & -0.067 & 0 & 0.564 & 1 & 0.564&0 & -0.067 \\
\hline
m+1 & -0.043j & -0.125 & 0.206j & 0.239 & -0.206j &-0.125 & 0.043j \\
\hline
\end{tabular}
\end{table}
\section{Simulation Result}
\subsection{Power boosting performance}
\begin{figure}[htbp]
\centering
\includegraphics[width=8cm]{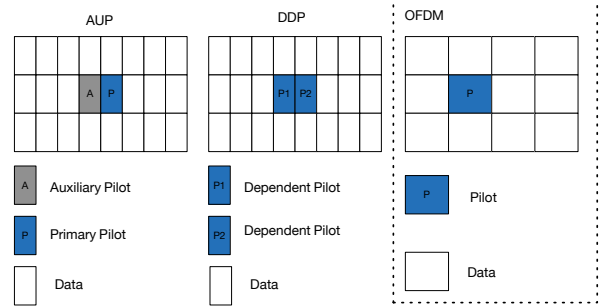}
\centering
\caption{A data block for FBMC/OQAM and OFDM}\label{RS_Structure}
\end{figure}
As discussed in section \ref{Perf_Com}, both schemes require power boosting when a desired SNR for clean pilot is to be achieved. In order to verify this power boosting reduction of the DDP scheme, the simulation is performed in an AWGN channel so that the effect of the time varying multi-path channel could be eliminated. A data block is set up as shown in Fig.\ref{RS_Structure} where pilots are surrounded by data symbols generating intrinsic interference. An OFDM system is included as a benchmark. As OFDM only has halved symbol rate compared to FBMC/OQAM, only one complex pilot is inserted for OFDM wherever two PAM pilots are inserted for FBMC/OQAM system. The data of FBMC/OQAM applies PAM with 2 constellations which have power of $\rho^{2}=1/2$ while the data of OFDM applies QPSK with unit power. The power of the complex noise is defined as $\sigma^{2}=Var(n_{m,n}) \forall (m,n)$. As the data only suffers real part of the complex noise at the FBMC/OQAM receiver, thus the SNR of the data for both systems is the same: $SNR_{Data}=1/\sigma^{2}$. The power setting for pilot is to ensure that the clean pilot has the same SNR as the data, i.e. $SNR_{AUP}=SNR_{DDP}=SNR_{OFDM}=SNR_{Data}$. The PHYDYAS filter is used with overlapping factor $K=4$ \cite{PHYDYAS}.

\begin{figure}[htbp]
\centering
\includegraphics[width=8cm]{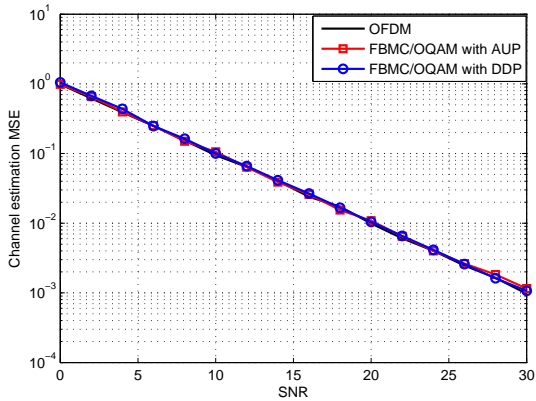}
\centering
\caption{Channel estimation performance}\label{MSE_AWGN}
\end{figure}

Fig.\ref{MSE_AWGN} shows the mean square error (MSE) of the channel estimation performance for different schemes. As the target SNR for the clean pilot is normalized for all schemes, the MSE performances align to each other as anticipated. In such case, the transmission power determines the overall channel estimation efficiency. As there is no power boosting in OFDM system, we define the power boosting ratio as $\eta_{AUP}=\frac{P_{AUP}}{P_{OFDM}}$ and $\eta_{DDP}=\frac{P_{DDP}}{P_{OFDM}}$ where $\eta_{OFDM}=1$. The comparison of power boosting ratio is shown in Table \ref{Power_Boosting}. From the table, the power boosting of DDP is significantly reduced comparing with AUP.
\begin{table}
\centering
 \caption{Power boosting ratio}\label{Power_Boosting}
\begin{tabular}{|c|c|c|c|}
\hline
 & OFDM & FBMC/OQAM with AUP &  FBMC/OQAM with DDP \\
\hline
 $\eta$ & 1 & 2.01 &  1.49\\
 \hline
\end{tabular}
\end{table}
\subsection{Performance in Practical Systems}
\begin{figure}[htbp]
\centering
\includegraphics[width=8cm]{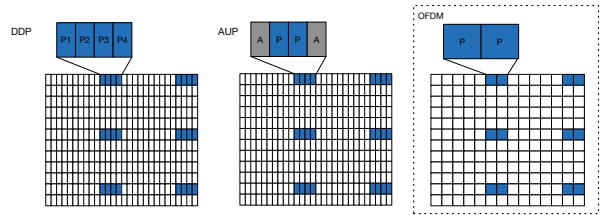}
\centering
\caption{Reference signal structure based on LTE downlink system}\label{LTE_RS}
\end{figure}

\begin{figure}[htbp]
\centering
\includegraphics[width=8cm]{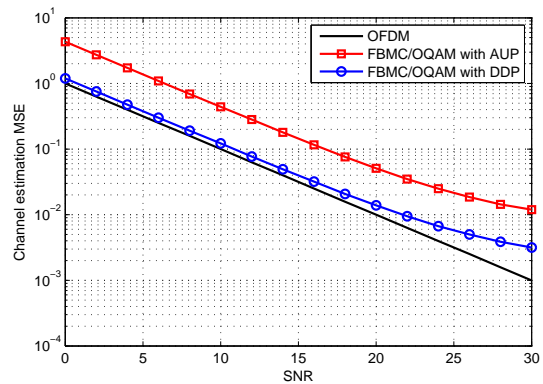}
\centering
\caption{Channel estimation performance with normalized transmission power}\label{MSE_DMRS}
\end{figure}

\begin{figure}[htbp]
\centering
\includegraphics[width=8cm]{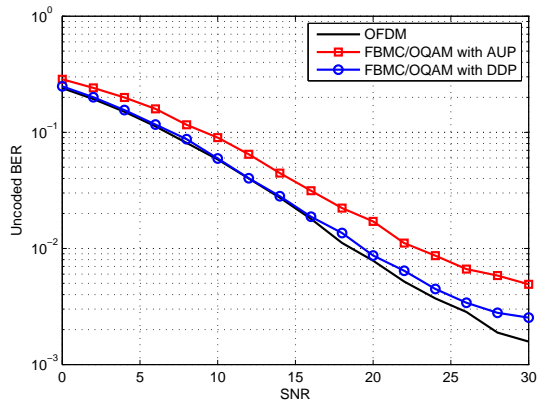}
\centering
\caption{BER performance with normalized transmission power}\label{BER}
\end{figure}

The previous subsection verifies the power boosting reduction of the DDP, which agrees the theoretical analysis. However, the performance of these schemes should be evaluated in a practical system where pilot design must consider more issues: firstly, a practical transmitter normally has power limitation, which is due to the power amplifier's limitation or interference concerns. This means the pilot's power can't be boosted freely; secondly, a practical system always requires multiple pilots with specially designed structure. Thus, we select the practical LTE downlink system as an example for evaluation \cite{LTE}. A data block in LTE contains 14 OFDM symbols with multiple pilots inserted. The pilots are organized as in Fig.\ref{LTE_RS}. Two consecutive pilots are grouped as a pair and multiple pairs are scattered over the block. For the sake of analogy, the FBMC/OQAM is designed with 28 symbols, four consecutive pilots are grouped together and the same number of groups are scattered over the block. The transmission power of each scheme is normalized for each scheme, i.e. $P_{AUP}=P_{DDP}=1$ for each two pilots. Consequentially, the effective power of the clean pilot is different for each scheme. For AUP, the central two subcarriers are reserved for primary pilots, which are used by the receiver for channel estimation. Each auxiliary pilot is to neutralize the interference of the adjacent primary pilots. For DDP, four dependent pilots are jointly calculated following the interference neutralization principle. Note the solution is no longer the same as \ref{DDP}. Two dependent pilots are used by the receiver to generate one complex clean pilot and thus two complex clean pilots in total are obtained by the receiver. The ETU channel with 50km/h speed is generated and each FBMC/OQAM symbol is convoluted with the time varying channel before combination at the receiver.
The channel estimation MSE is shown in Fig.\ref{MSE_DMRS}. The DDP scheme has almost no loss in most SNR ranges while AUP scheme has a very big loss at all SNR points. With high SNR, the frequency-selectivity and the time variance of the channel break the flat channel assumption and thus lead to degradation for FBMC/OQAM system. Approximately, 6dB gain could be achieved by DDP over the AUP. Such improvement is due to the efficient interference utilization property of DDP. When 4 consecutive pilots are grouped together, DDP could perfectly utilize the interference between each pilot and thus reduces the power boosting requirement to an extreme level. On contrary, AUP still has to spend a lot of power on interference neutralization, which leaves very small power for the clean pilot when power limitation is imposed.
Fig. \ref{BER} shows the uncoded BER performance with each channel estimation scheme. The receiver estimates the frequency domain channel response at the pilot subcarriers and then performs interpolations to obtain the channel estimates at data subcarriers. Equalization is done based on the channel estimates before hard decision is made. As shown, the FBMC/OQAM with DDP scheme has the same BER performance as OFDM except for very high SNR points. Such loss is due to the intrinsic interference of FBMC/OQAM in frequency selective channel rather than channel estimation errors. FBMC/OQAM with AUP scheme has degraded BER performance, which is mainly because of the poor channel estimation performance.

\section{Conclusion}
In this paper, we discuss the scattered pilot design for FBMC/OQAM systems. The paper firstly gives the general solution to the interference neutralization problem and demonstrates that existing schemes are only specific solutions to the general solution. By breaking the primary pilot principle adopted by existing schemes, the paper shows another specific solution, which relies on no independent pilot, can significantly improve the performance. At last, the simulation results clearly show the new scheme could achieve close-to-OFDM performance in FBMC/OQAM system even with the presence of intrinsic interference.





%\bibliographystyle{IEEEtran}

\end{document}